\documentstyle[epsfig]{mn} 
\newcommand{\etalc}{et~al.}
\newcommand{\degsym}{$^{\circ}$}

\begin{document}

\title[Variability of the Accretion Stream in the Eclipsing Polar EP Dra]
{Variability of the Accretion Stream in the Eclipsing Polar EP Dra}

\author[C. M. Bridge et al.]  {C.M. Bridge$^1$, Mark Cropper$^1$, Gavin
Ramsay$^1$, J.H.J de Bruijne$^2$,\and A.P. Reynolds$^2$, M.A.C. Perryman$^2$\\
$^1$ Mullard Space Science Laboratory, University College London, Holmbury
St. Mary, Dorking, Surrey, RH5 6NT\\ 
$^2$ Research and Scientific Support Department of ESA, ESTEC, Postbus 299, 2200 AG Noordwijk, The Netherlands}

\maketitle

\begin{abstract}

We present the first high time resolution light curves for six eclipses of the
magnetic cataclysmic variable EP~Dra, taken using the superconducting tunnel
junction imager S-Cam2. The system shows a varying eclipse profile between
consecutive eclipses over the two nights of observation. We attribute the
variable stream eclipse after accretion region ingress to a variation in the
amount and location of bright material in the accretion stream. This material
creates an accretion curtain as it is threaded by many field lines along the
accretion stream trajectory. We identify this as the cause of absorption
evident in the light curves when the system is in a high accretion state. We do
not see direct evidence in the light curves for an accretion spot on the white
dwarf; however, the variation of the stream brightness with the brightness of
the rapid decline in flux at eclipse ingress indicates the presence of some
form of accretion region. This accretion region is most likely located at high
colatitude on the white dwarf surface, forming an arc shape at the foot points
of the many field lines channeling the accretion curtain.

\end{abstract}


\section{Introduction}

Polars are a sub-class of magnetic cataclysmic variables. They are binary
systems containing a strongly magnetic, accreting white dwarf primary
(10$\--$200~MG) and a late main sequence secondary. The magnetic field of the
primary controls the flow of material lost from the secondary and prevents the
formation of an accretion disk. The material is instead confined by the
magnetic field and channeled to accrete directly on to the primary through a
stand-off shock (see Cropper 1990 for a review of polars).

The eclipsing nature of a small subset of these systems allows the isolation of
the emission from different regions as various parts are eclipsed and revealed
by the secondary. In particular these systems are well suited to the study of
the brightness distribution of the accretion stream itself, as the emission
from this part of the system can be isolated once the accretion region (or
regions) has been eclipsed.

In this paper we present the first high time resolution observations of the
eclipses in the polar EP~Dra using a detector with intrinsic time and energy
resolution. The system has an orbital period of 104.6~minutes, and was
identified as the optical counterpart to a hard X-ray source in the {\sl
HEAO~1} survey by Remillard~\etalc\ (1991). The system is faint, with an
average faint phase brightness of V$\simeq$18 in 1992 and 1995 and a maximum
brightness of V$\simeq$17 (Schwope \& Mengel 1997; hereafter SM97), with
evidence for accretion at one region only on the white dwarf primary (Remillard
\etalc\ 1991).

Our observations were taken with the second prototype of a new generation of
optical detector (S-Cam2) which uses superconducting tunnel junctions (STJs) to
record the time of arrival, location on the array, and, uniquely, the energy of
incident photons. For details of the instrumentation see e.g. Rando~\etalc\
(2000). Previous observations of close binary systems made using S-Cam2 include
UZ For (Perryman~\etalc\ 2001), HU Aqr (Bridge \etalc\ 2002) and IY UMa
(Steeghs~\etalc\ 2002).

\section{Observations and reductions}
\label{sec:observations}

\begin{table*}
\caption{Summary of observations of EP Dra for the nights of 2000 October 2/3
and 2000 October 3/4. Cycle numbers are with respect to the ephemeris of
SM97.}
\begin{center}
\begin{tabular}{clrrrrr}
\hline
Cycle   & Date	& Start time 	& Observation 	& Time of mid-eclipse & Time of
mid-ingress & Time of mid-egress\\
number  &	& (TDB)      	& length (s)  	& (TDB) & (TDB)& (TDB)\\
(56900+)&	&(2450000.0+)	&		&(2450000.0+)&(2450000.0+)&(2450000.0+)\\
\hline
62  & 2000 Oct 2 & 1820.36254 & 1496 & 1820.37575 & 1820.37382& 1820.37850\\
63  & 2000 Oct 2 & 1820.44230 & 1800 & 1820.44840 & 1820.44649& 1820.45128\\
64  & 2000 Oct 3 & 1820.50681 & 2523 & 1820.52106 & 1820.51913& 1820.52387\\
76  & 2000 Oct 3 & 1821.38171 & 2400 & 1821.39294 & 1821.39097& 1821.39572\\
77  & 2000 Oct 3 & 1821.45653 & 2111 & 1821.46559 & 1821.46368& 1821.46835\\
78  & 2000 Oct 4 & 1821.53007 & 1393 & 1821.53825 & 1821.53631& 1821.54112\\
\hline
\end{tabular}
\end{center}
\label{tab:observations}
\end{table*}

The observations were made at the William Herschel Telescope, La Palma, on the
nights of 2000 October 2/3 and 2000 October 3/4. The S-Cam2 instrument was
located at the Nasmyth focus and a total of six eclipses of EP~Dra were
obtained. Table~\ref{tab:observations} gives the cycle number (relative to the
ephemeris of SM97), date, start time and length for each observation, as well
as the time of mid-eclipse (defined as orbital phase $\phi=0.0$), mid-ingress
and mid-egress of the steep components of the eclipse (all times are in TDB,
which includes light travel time correction to the solar system
barycentre). Two of the observations (cycles 62 and 64) have data gaps caused
by the instrument exceeding the data acquisition limits.

The useful observational wavelength range of S-Cam2 is 340$\--$680~nm
(Perryman~\etalc\ 2001). We split this wavelength range into four bands, which
we label here as U (340$\--$400~nm), B (390$\--$490~nm), V (500$\--$600~nm) and
R$_{c}$ (590$\--$680~nm) and which correspond very broadly to those of the
Johnson-Cousins UBVR$_{c}$ system (Bessell 1990, and references therein). The
light curves were background subtracted using a mean value from off-source
pixels taken from the mid-eclipse phases. See Perryman~\etalc\ (2001) for a
detailed discussion of the data reduction process.

\begin{figure*}
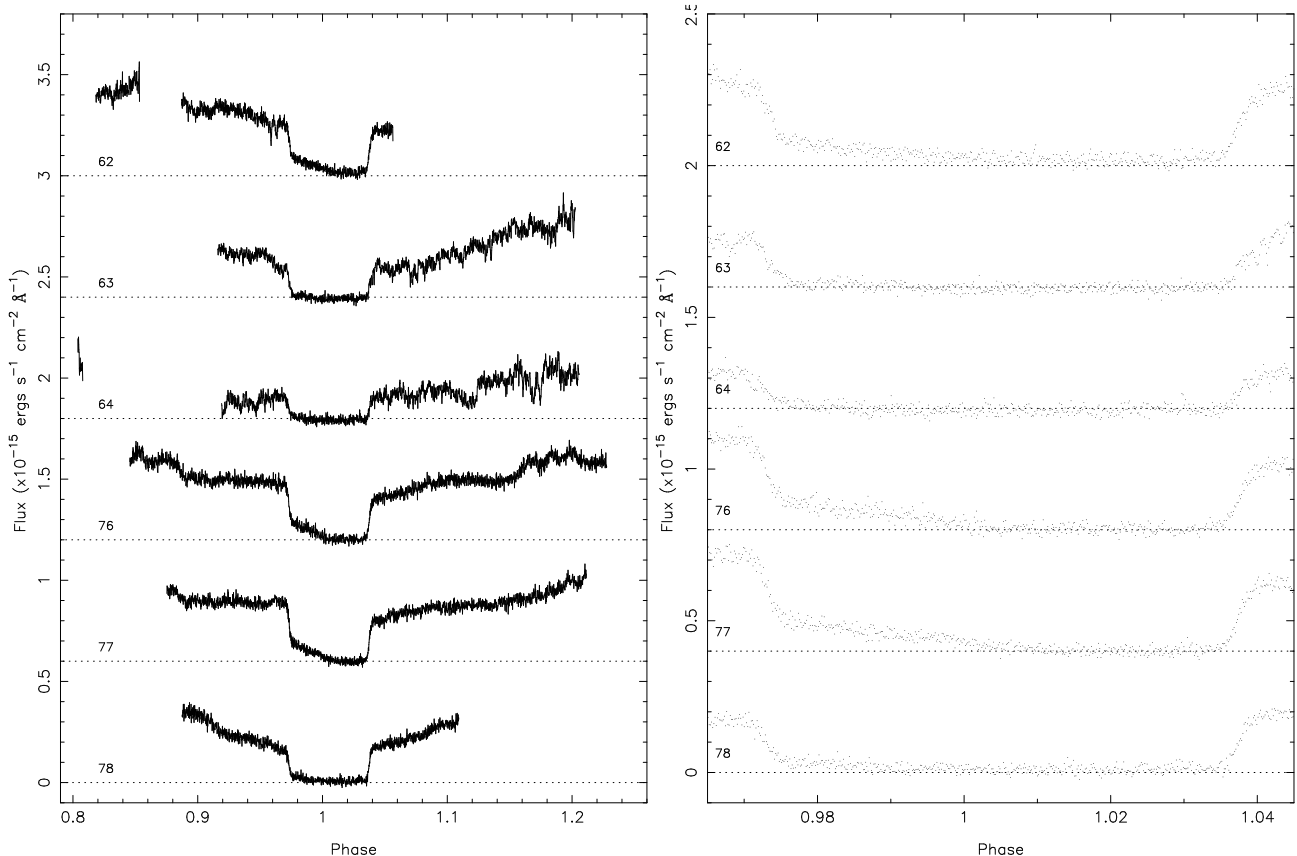

\epsfig{file=./fig1a.ps,width=8.5cm,angle=0}
\epsfig{file=./fig1b.ps,width=8.5cm,angle=0}
\caption{White light curves for EP Dra in 1~s time bins. Each cycle is offset
vertically by 0.6$\times$10$^{-15}$ ergs s$^{-1}$ cm$^{-2}$ \AA$^{-1}$, and
phased according to the ephemeris of SM97. The right hand panel shows the
eclipse of the accretion region and stream in more detail, highlighting the
variability between consecutive eclipses. Each cycle is offset vertically by
0.4$\times$10$^{-15}$ ergs s$^{-1}$ cm$^{-2}$ \AA$^{-1}$.}
\label{fig:whitelight}
\end{figure*}

\begin{figure*}
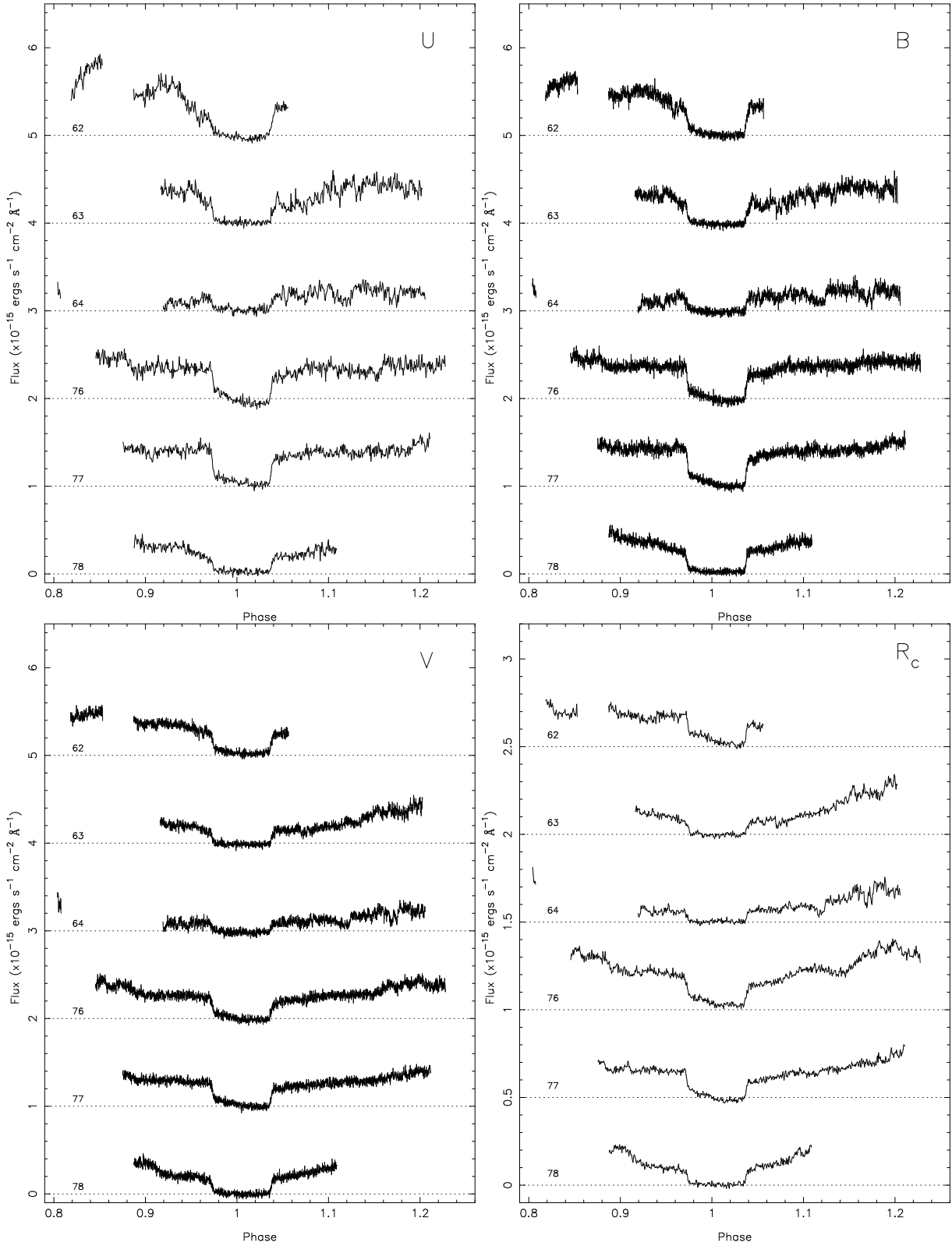

\epsfig{file=./fig2a.ps,width=8.cm,angle=0}
\epsfig{file=./fig2b.ps,width=8.cm,angle=0}
\vspace{1.0mm}
\epsfig{file=./fig2c.ps,width=8.cm,angle=0}
\epsfig{file=./fig2d.ps,width=8.cm,angle=0}
\caption{UBVR$_{c}$ colour light curves for the six cycles, phased with respect
to the ephemeris of SM97. The U and R$_{c}$ band light curves are binned in 4~s
time bins due to low count rates. The B and V bands are in 1~s bins. Note the
vertical scale for the R$_{c}$-band is smaller due to the lower flux in this
band.}
\label{fig:UBVRcurves}
\end{figure*}

The `white' light curves are shown in Figure~\ref{fig:whitelight} and
UBVR$_{c}$-bands in Figure~\ref{fig:UBVRcurves}. Each figure shows the three
consecutive eclipses for each of the two nights of the observations. The right
hand panel of Figure~\ref{fig:whitelight} shows the eclipse of the white dwarf,
accretion region and accretion stream on an expanded scale, making the rapid
changes at ingress and egress more clear. The U and R$_{c}$ light curves are
binned in 4~s intervals and the B and V in 1~s. The light curves have been
calibrated in energy units using observations of the standard star BD+28\,4211,
which were taken on the first night. A standard star was not observed on the
second night so cycles 76 to 78 are only approximately calibrated. The maximum
brightness is V $\approx$ 17 at $\phi$ = 1.2, consistent with the maximum
measured by SM97.

The seeing for the first night was $\sim$1.5 arcsec, and for the second night
$\sim$1 arcsec. Some of the variability in the eclipses of the first night may
be in part due to the occasionally poorer seeing, causing the point spread
function to spill over the edges of the 6x6 array. A number of sharp dip
features (e.g. at $\phi\sim$ 0.96 in cycle 62) can be attributed to this.

\section{Light curve features}
\label{sec:lcs}

\subsection{White dwarf and accretion region eclipse}
\label{sec:wdeclipse}

The most prominent feature of the light curves is the eclipse of the white
dwarf and accretion stream by the secondary between $\phi$ = 0.972 and $\phi$ =
1.04. The duration of the eclipse (measured between the middle of the steep
eclipse components) is the same for all cycles, 6.8~minutes, equivalent to
$\Delta\phi$ = 0.0646.

\begin{figure}
\centerline{\epsfig{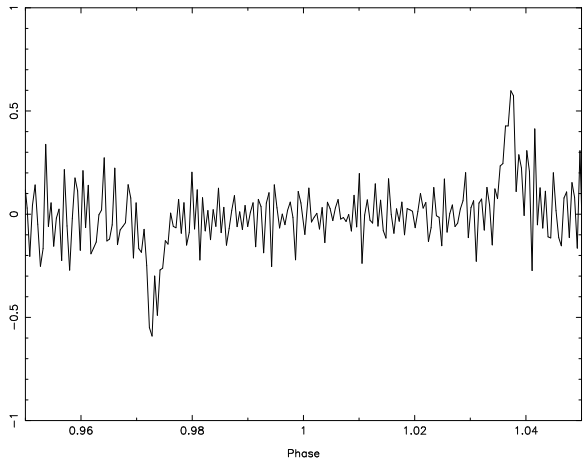}}
\caption{The mean of the six 3~s differential white light curves.}
\label{fig:diff}
\end{figure}

The eclipse of the white dwarf photosphere and the accretion region is very
rapid. To characterise the rapid eclipse we binned the six white light curves
into 3~s time bins, then for each time bin we subtracted the brightness of the
previous time bin. Figure~\ref{fig:diff} shows the mean of these six
differential light curves. The width of the complete ingress and egress of the
white dwarf and accretion region is 36~s which is comparable to the expected
ingress duration for the white dwarf of 37~s (assuming a primary mass
$M_{1}=0.43~M_{\odot}$ and mass ratio $q=0.31$; SM97). The amplitude of the
brightness change at ingress is greatest at bluer wavelengths, which is as
expected for a hot accretion region and white dwarf.

HU Aqr (Bridge~\etalc\ 2002) and UZ~For (Perryman~\etalc\ 2001) show a rapid
drop in flux at eclipse ingress lasting a few seconds, which is taken as
evidence of a bright, compact accretion region. In contrast we see no evidence
for such a rapid drop in EP Dra. However, there is some asymmetry in
Figure~\ref{fig:diff} which implies the presence of an accretion region in some
form.

\begin{figure}
\centerline{\epsfig{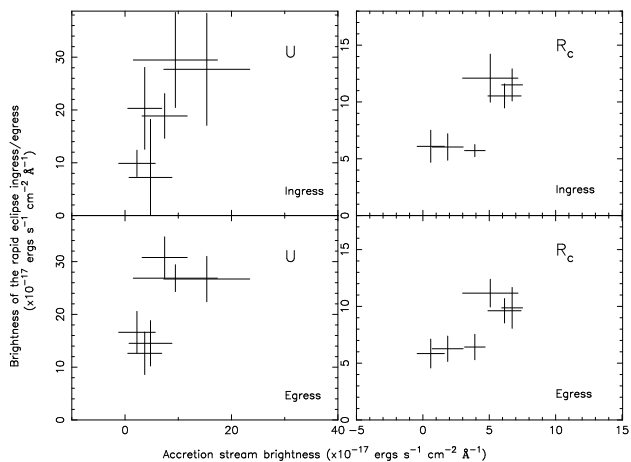}}
\caption{The amplitude of the rapid brightness change at ingress and egress
against the brightness of the accretion stream. This is shown for both the U
and R$_{c}$ band light curves. The stream brightness is measured immediately
after the steep eclipse ingress at $\phi$ = 0.975.}
\label{fig:spotstream}
\end{figure}

Following the rapid 36~s decline in brightness seen at ingress, the stream is
the major contributor to the observed brightness, with only a small amount of
emission from the secondary in the red. The presence of an accretion region is
also evident from our observation that the amplitude of the rapid ingress
varies with the brightness of the accretion stream, and this is illustrated in
Figure~\ref{fig:spotstream}. If there was no contribution from an accretion
region, then we would expect no correlation. We therefore identify the rapid
decline in flux with the eclipse of both the white dwarf and an accretion
region, despite the accretion region not being directly evident in the
eclipse. By comparison with HU~Aqr and UZ~For, the absence of a compact
accretion region is surprising.

\subsection{Stream eclipse}
\label{sec:streameclipse}

In our light curves the stream eclipse ingress, from $\phi\simeq$~0.975
onwards, takes two forms: a long ingress from a relatively bright stream
(cycles 62, 76 and 77) and a short ingress from a faint stream (cycles 63, 64
and 78). Figure~\ref{fig:smoothplot} shows the eclipse profile of all six
cycles, and the two forms of stream eclipse are clear. We also see that there
is a decline in flux preceding the ingress of the white dwarf and accretion
region in cycles 63 and 78 where the stream is relatively faint.

\subsection{Out of  eclipse}
\label{sec:preposteclipse}

The prominent feature before and after the eclipse is the decline in brightness
at phase $\phi$ = 0.874 in cycle 76 and subsequent rise at $\phi$ = 1.16. This
is seen in all bands in Figure~\ref{fig:UBVRcurves}. The duration of the
decline is $\sim$ 100~s, ending at $\phi$ = 0.89. The reduced phase coverage in
cycle 77 means that we cannot be certain that the feature at $\phi$=0.88 is the
same feature as that in the preceding cycle. However, the similarity in the
shape of the light curve compared to that of cycle 76 and the location of the
feature means that we proceed assuming they are the same. We refer to this
pre-eclipse decline and post-eclipse rise as the `trough'.

There is a bright stream in cycle 76; however in cycle 78 the stream is fainter
and we identify the trough as starting at the later phase of $\phi$ = 0.90. We
also infer that the rise at $\phi$ = 1.08 in cycle 78 is the counterpart of
that at $\phi$ = 1.16 in cycle 76.

Following the start of the trough, cycles 76 and 77 show a relatively flat
light curve prior to the eclipse of the white dwarf. In cycles 62, 63 and 78,
after $\phi$=0.95, there is a decline in flux prior to the eclipse ingress.

After the rapid egress of the white dwarf photosphere and accretion region at
$\phi=1.04$, there is no obvious stream egress in cycles 63 and 64,
corresponding to a lack of a stream brightness at ingress. In cycles 76, 77 and
78 the accretion stream egress is seen as a gradual increase in flux, followed
by a relatively flat light curve in cycles 76 and 77.

\section{Discussion}
\label{sec:discussion}

\subsection{Accretion region location}
\label{sec:accgeo}

In determining the location of material confined to magnetic field lines, the
values chosen for $R_{\mu}$ (the distance from the white dwarf at which
material is threaded by the field lines), and the colatitude (angle measured
from the spin axis of the white dwarf) and longitude of the accretion region
are important. Two different estimates for the location of this region are
proposed by Remillard~\etalc\ (1991) and SM97. Based primarily on their
circular polarimetry data, Remillard~\etalc\ (1991) concluded that there was
only one accretion region, at a colatitude of $\sim$ 18\degsym and longitude
$\sim \--$17\degsym. On the other hand, SM97, modeling the bright cyclotron
peaks in optical observations, found a colatitude of $\sim$ 60\degsym\ $\--$
70\degsym.

As we are unable to say with certainty what the true location of the accretion
region is, we consider both possibilities of high and low latitude accretion in
the following discussion. To proceed we use the mass ratio ($q = 0.31$; SM97)
to create a model Roche lobe and white dwarf with a magnetic pole at a given
colatitude, $\beta$, and longitude, $\zeta$. Dipole field lines are then
constructed passing through a given $R_{\mu}$. By varying $R_{\mu}$ for a given
magnetic pole location, field lines can be created that eclipse a given part of
the polar system at a phase given by features observed in the light
curves. Although Remillard~\etalc\ (1991) and SM97 quote the location of the
accretion region, we use these values for the location of the magnetic pole,
noting that this will cause accretion at a region offset from these values for
a given $R_{\mu}$.

\begin{figure}
\centerline{\epsfig{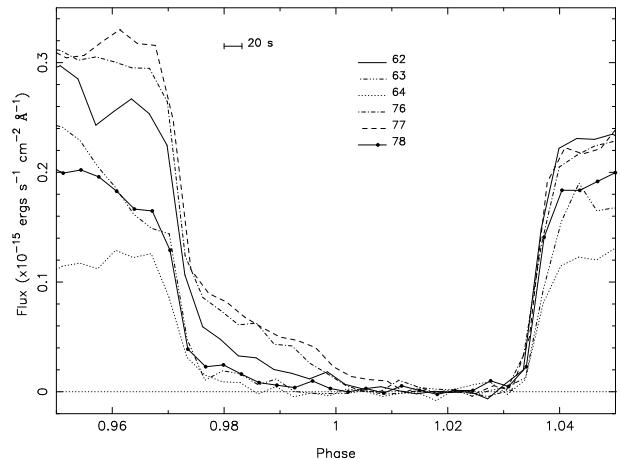}}
\caption{The six white light curves centred on the eclipse by the
secondary. They have been binned into 20~s, which removes any features related
to the eclipse of the white dwarf and the accretion region, and the secondary
contribution has been subtracted.}
\label{fig:smoothplot}
\end{figure}

\subsection{Stream variations}
\label{sec:streamvariations}

The observed stream variations (Figures~\ref{fig:whitelight} and
\ref{fig:smoothplot}) are essentially caused by a change in the visibility of
material confined to the magnetic field lines of the white dwarf. This could be
caused by an asynchronism in the white dwarf spin and orbital periods, causing
a changing magnetic field to be presented to the incoming accretion
stream. However, we consider this unlikely due to a lack of a change in the
relative position of the bright phase and eclipse between the observations of
Remillard~\etalc\ (1991) and SM97.

Alternatively, there is either a change in the amount of bright material on
visible field lines or material is located on different field lines. These can
be caused by a change in $R_{\mu}$, from a variation in the mass transfer rate
and an inhomogeneity in the flow. It is not possible from our data to isolate
either possibility as the cause, and it is probable that both contribute to the
observed stream variability. Variations in the accretion stream eclipse profile
have been observed in many eclipsing polars, for example in HU Aqr
(Glenn~\etalc\ 1994; Harrop-Allin~\etalc\ 1999; Bridge~\etalc\ 2002) and
V895~Cen (Salvi~\etalc\ 2002).

We can estimate $R_{\mu}$ from the phase at which visible magnetic field lines
are completely eclipsed in our geometric model. We find that a longer stream
ingress requires a smaller $R_{\mu}$ for a high colatitude accretion region
(SM97), compared with that from a low colatitude (Remillard~\etalc\ 1991).

\begin{figure}
\centerline{\epsfig{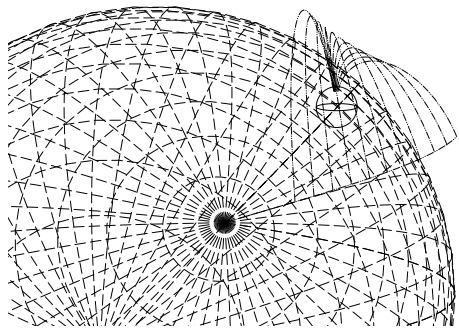}}
\caption{Plot of the geometry of EP Dra consistent with a variable brightness
stream for $\phi$ = 0.98. The magnetic field lines are still visible, causing a
long stream ingress in the light curves for bright stream material confined to
these field lines. The magnetic pole is located at $\beta$ = 18.0\degsym\ and
$\zeta$ = --17.0\degsym (assuming the values for the accretion region in
Remillard~\etalc\ 1991 are the same as the magnetic pole).}
\label{fig:binvis}
\end{figure}

\subsection{Trough feature}
\label{sec:trough}

We consider two mechanisms that could explain the trough: absorption or
emission. Absorption would occur from material lost by the secondary, with the
initial drop in the light curves resulting from the onset of the
absorption. Alternatively, the feature could be caused by the changing
cyclotron emission with phase.

\subsubsection{Emission}
\label{sec:emission}

\begin{figure}
\centerline{\epsfig{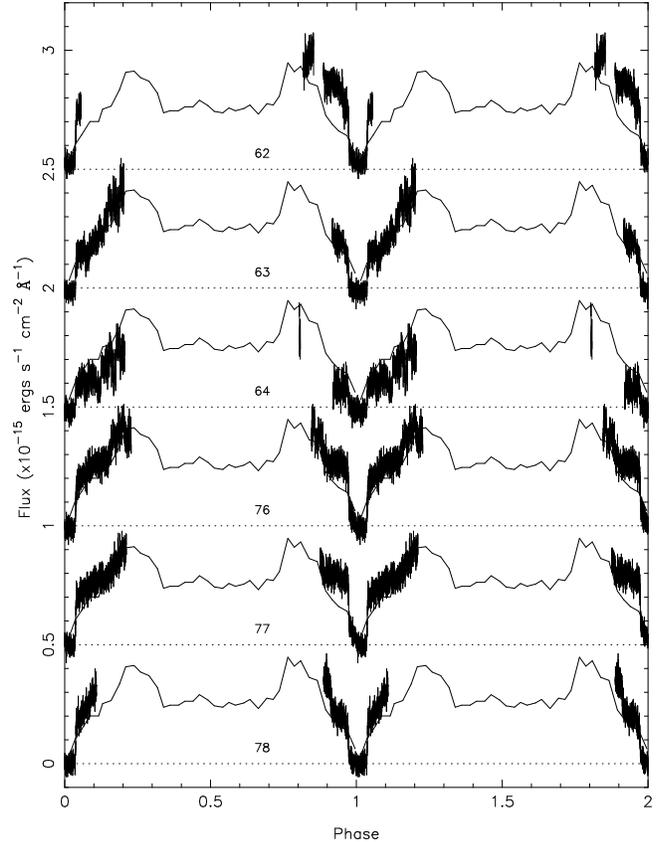}}
\caption{The V-band light curves of EP~Dra plotted on the same phase range as
those of SM97 (their figure 1). Over plotted as a solid line is their 1995
V-band observation for comparison.}
\label{fig:overlaid}
\end{figure}

SM97 observed a sharp rise to and decline from the bright phase (which is
indicative of the cyclotron beaming effect) in their V-band light curves of
EP~Dra. The phase range of our light curves is much less than that of SM97 so
we do not observe the same behaviour either side of the eclipse; however, the
trend of our light curves is consistent with such a rise to and fall from
maximum. Figure~\ref{fig:overlaid} shows our V-band light curves with the SM97
figure 1 V-band data from 1995. The SM97 data has been scaled vertically to
show the coincidence of the features.

The wavelength at which the cyclotron emission peaks is dependent upon the
magnetic field strength of the white dwarf, with increasing field strength
shifting the peak to shorter wavelengths. For EP~Dra, the field is $\sim$16~MG
(SM97) so we would expect the orbital variability to be greatest towards redder
wavelengths. We therefore expect to see a significant change in the shape of
the light curves in the different colour bands, and hence a prominent feature
in the colour ratios. As a signature of the cyclotron emission, we would expect
a sharp peak in the R$_{c}$-band followed by a gradual decline, and conversely
in the U-band a more gradual rise in flux to a maximum at $\phi\approx$ 0.97,
at which point the accretion region is eclipsed, as for example in WW~Hor
(Bailey~\etalc\ 1988; their figure 1).

Thus while there is evidently some cyclotron beaming, the absence of any
significant features in the light curves or consistent variations in the colour
ratios implies that cyclotron beaming is only partially responsible for the
observed light curve shape, and cannot explain the flat light curve pre and
post eclipse.

\subsubsection{Absorption}

A second possible cause of the trough is absorption by obscuring material lost
from the secondary in an accretion flow. This would be supported by the
correlation between the presence of this feature and the brightness of the
stream (Section~\ref{sec:preposteclipse}), and also changes in $R_{\mu}$ caused
by differing amounts of material in the stream
(Section~\ref{sec:streamvariations}).

We see a reddening in the U/R$_{c}$ colour ratios during the decline in flux in
cycle 76 between $\phi\approx$ 0.88 and 0.89. This effect is more pronounced in
cycle 78 where there is a significant dip towards the red and rise to the blue
again during the extended decline between $\phi\approx$ 0.89 and 0.925. In the
U and B-bands of cycle 62 we see a rise in flux and a corresponding decrease in
the R$_{c}$-band at phases corresponding to the end of the extended decline in
cycle 76. The most likely cause is the onset of absorption by material in the
accretion stream. The increased absorption at shorter wavelengths is contrary
to the expected free-free absorption in the accretion stream (King \& Williams
1985), which predicts increasing absorption at longer wavelengths
(Watson~\etalc\ 1995).

The observed variation of the light curves with the colour ratios suggests that
the absorption process is likely to be bound-free absorption of the Balmer
continuum. This is predominantly in the U-band, and the variations in the
colour ratios and light curves are explained by variations in the amount and
density of material confined by the magnetic field lines of the white dwarf.

\subsection{The accretion flow}

The phase of the onset of the trough is later than that of the pre-eclipse dip
seen in the light curves of HU Aqr (Harrop-Allin \etalc\ 1999; Bridge \etalc\
2002), which was identified as the eclipse of the accretion region by the
strongly collimated accretion stream. This phase is determined by the geometry
of the field lines carrying the accreting material, so a difference between
systems is not unexpected. While the onset of the trough feature is consistent
with an eclipse caused by the accretion stream, an extended accretion curtain
would cause absorption over an extended phase range (previous section), and
produce the observed flat light curves during the trough
(Figures~\ref{fig:whitelight} and \ref{fig:UBVRcurves}). The extent of this
accretion curtain can be estimated using the eclipse light curve features and
the geometric model introduced earlier. 

A value of $R_{\mu}$ can be estimated by varying the parameter until the field
lines begin to eclipse the accretion region, at the phase set by the start of
the trough in the light curve. For those cycles with brighter streams and
accretion regions, where the onset of the trough is earlier, we find values of
$R_{\mu}\sim0.14a$ (for $\beta=65^{\circ})$ and $0.19a$ (for
$\beta=18^{\circ}$). For those cycles where the onset of the trough is later at
$\phi=0.90$, we find $R_{\mu}\sim0.16a$ (for $\beta=65^{\circ}$) and $0.22a$
(for $\beta=18^{\circ}$). The outer edge of the curtain, where material threads
closest to the secondary, can be estimated from the end of the egress of bright
stream material around $\phi$ = 1.1, and is independent of $\beta$. This
assumes there is no significant continuum emission from the ballistic section
of the accretion stream, where material is expected to be faint and cooling as
it falls. For those cycles where the bright stream egress is clearly seen, we
find a value of $\sim0.42a$. For those cycles with fainter streams the egress
of the stream is not seen. We therefore infer that brighter streams result in a
wider accretion curtain, with more material in the accretion stream penetrating
further into the magnetosphere.

For faint stream cycles a decline in flux is seen immediately before the
eclipse of the accretion region (see Section~\ref{sec:streameclipse}). This
could be due to the eclipse of stream material which is threading early in the
trajectory, and which should emit through magnetic heating. Alternatively the
decline could be caused by absorption by material along the line of sight to
the white dwarf and accretion region (see Section~\ref{sec:trough}).

The trough is absent in the observations of Remillard~\etalc\ (1991) and
SM97. The stream also appears to be fainter in both previous observations,
although this may be due to the poorer sampling. The lack of a bright stream
may be the cause of this absence of a trough in their light curves, if the two
are linked as we suggest.

An extended accretion curtain with material being threaded at different
$R_{\mu}$ would allow material to accrete over an extended region on the white
dwarf. The foot points of the field lines form an arc shape on the white dwarf
surface. Such an extended accretion region would be consistent with the optical
cyclotron models of SM97 which show evidence for an accretion arc or ribbon.

\section{Conclusions}

We have analysed the first high signal-to-noise ratio and high time resolution
data of EP~Dra taken on two consecutive nights at the WHT using S-Cam2. The
eclipse light curves show variability in the accretion stream and accretion
region over the timescale of the orbital period. We see no direct evidence in
the light curves for the expected rapid eclipse of a small accretion region on
the white dwarf. The rapid eclipse seen in the light curves is a combination of
emission from the white dwarf photosphere and the accretion region. We see
evidence for the variability of the accretion region from the variation in
brightness of the rapid eclipse ingress with the varying accretion stream
brightness.

Variability seen in the light curves on a longer timescale is influenced to
some extent by cyclotron beaming. However from the colour dependence, there is
probably also a contribution from absorption, and this is seen as a trough in
the light curves of the second night. We attribute the absorption to bound-free
absorption by material in an extended accretion curtain obscuring the accretion
region and white dwarf. There may also be significant absorption by material
located close to the white dwarf above the accretion region.

Accreting material is threaded onto many field lines along the accretion stream
trajectory, and the location in phase of the onset of the trough or absorption
dip provides an estimate of the location of the edge of the accretion
curtain. Variations in the brightness of the accretion stream seen after the
ingress of the white dwarf and the accretion region are caused by a change in
the location of bright stream material in the accretion curtain and/or a change
in the extent of the curtain. From the extent of the accretion curtain we infer
the presence of an extended accretion arc at the foot points of the accreting
field lines, however this region is still small compared with the size of the
white dwarf.

\section*{ACKNOWLEDGMENTS}

We acknowledge the contributions of other members of the Research and
Scientific Support Department of the European Space Agency at ESTEC involved in
the optical STJ development effort, in particular S.~Andersson, D.~Martin,
J.~Page, P.~Verhoeve, J.~Verveer, A.~Peacock and N.~Rando. We acknowledge the
excellent support given to the instrument's operation at the WHT by the ING
staff, in particular P.~Moore and C.R.~Benn. We would also like to thank Pasi
Hakala for valuable discussions.

\end{document}